\documentclass[twocolumn, showpacs,preprintnumbers,amsmath,amssymb]{revtex4}

\usepackage{graphicx}
\usepackage{dcolumn}
\usepackage{bm}

\begin{document}

\preprint{APS/123-QED}

\title{Wakes in inhomogeneous plasmas}

\author{Roman Kompaneets}
\affiliation{Max-Planck-Institut f\"ur extraterrestrische Physik, 85741 Garching, Germany}

\author{Alexei V. Ivlev}
\affiliation{Max-Planck-Institut f\"ur extraterrestrische Physik, 85741 Garching, Germany}

\author{Vladimir Nosenko}
\affiliation{Max-Planck-Institut f\"ur extraterrestrische Physik, 85741 Garching, Germany}

\author{Gregor E. Morfill}
\affiliation{Max-Planck-Institut f\"ur extraterrestrische Physik, 85741 Garching, Germany}

\date{\today}

\begin{abstract}
The Debye shielding of a charge immersed in a 
flowing plasma is an old classic problem in plasma physics.
It has been given renewed attention in the last two decades in view of
experiments with complex plasmas, where charged dust particles are often levitated in a region with strong ion flow. 
Efforts to describe the shielding of the dust particles
in such conditions have been focused on the homogeneous plasma approximation, which
ignores the substantial inhomogeneity of the levitation region.
We address the role of the plasma inhomogeneity by rigorously calculating the point charge potential
in the collisionless Bohm sheath.
We demonstrate that the inhomogeneity can dramatically modify the wake,
making it non-oscillatory and weaker.
\end{abstract}

\pacs{52.27.Lw, 52.40.Kh, 52.30.-q, 52.35.Fp, 52.25.Mq}
\maketitle

\section{Introduction}
\label{introduction}
The Debye shielding of a charge immersed in a 
flowing plasma is an old classic problem,
which received considerable 
attention~\cite{neufeld-rit-phys.rev-1955, oppenheim-kam-phys.fluids-1964,
joyce-mon-phys.fluids-1967,
montgomery-joy-sug-plasma.phys-1968, cooper-phys.fluids-1969, sanmartin-lam-phys.fluids-1971,
laing-lam-fie-j.plasma.phys-1971, chenevier-dol-per-j.plasma.phys-1973,
stenflo-yu-shu-phys.fluids-1973, wang-joy-nic-j.plasma.phys-1981, peter-j.plasma.phys-1990, 
trofimovich-kra-sov.phys.jetp-1993, else-kom-vla-phys.rev.e-2010}, with applications ranging from 
charging of a spacecraft in the ionosphere~\cite{alpert-book, taylor-planet-space-sci-1967}
to spectra of ions moving through solids~\cite{bell-bet-pan-1976, jakubassa-1977}. 
Various forms of the potential distribution were obtained, depending on model assumptions
and parameter values.
For instance, in the collisionless case the far-field potential has been shown to generally scale as
$r^{-3}$~\cite{montgomery-joy-sug-plasma.phys-1968},
while in the presence of collisions it can have the $r^{-2}$-dependence~\cite{stenflo-yu-shu-phys.fluids-1973}. 
A flowing Maxwellian plasma can generate a series of potential wells downstream of the charge,
depending on the flow velocity~\cite{peter-j.plasma.phys-1990}.

The problem has been given renewed attention in the last two decades in view of
experiments with complex plasmas, where charged dust particles are often 
levitated in a region with strong ion flow (see, e.g., Refs.~\cite{fortov-ivl-khr-phys.rep-2005, ishihara-j.phys.d.appl.phys-2007,
morfill-ivl-rev.mod.phys-2009, shukla-eli-rev.mod.phys-2009, 
bonitz-hen-blo-rept.prog.phys-2010} for reviews of complex plasma research).
Much theoretical effort~\cite{benkadda-tsy-vla-1999, lampe-joy-gan-phys.plasmas-2000, lapenta-phys.rev.e-2000, 
schweigert-mel-pie-j.phys.iv.france-2000,
hou-wan-mis-phys.rev.e-2001, lampe-joy-gan-phys.scripta-2001, kompaneets-vla-ivl-new.j.phys-2008, kompaneets-kon-ivl-phys.plasmas-2007, ludwig-mil-kah-new.j.phys-2012,
block-car-lud-contrib.plasma.phys-2012} has been made to describe how the charged dust particles are shielded in that region,
as the shielding directly determines their mutual electrostatic interaction.
The suggested wake models range from those assuming cold flowing ions~\cite{vladimirov-nam-phys.rev.e-1995,
vladimirov-ish-phys.plasmas-1996, ishihara-vla-phys.plasmas-1997}
to advanced kinetic models incorporating ion-neutral collisions and
the electric field that drives the ion flow and supports the dust particles 
against gravity~\cite{schweigert-mel-pie-j.phys.iv.france-2000, kompaneets-kon-ivl-phys.plasmas-2007}.
A great deal of numerical simulations, based on various assumptions, have been performed~\cite{lapenta-phys.plasmas-1999, 
lapenta-phys.rev.e-2002,
miloch-vla-pec-phys.rev.e-2008,
miloch-tru-pec-phys.rev.e-2008,
miloch-plasma.phys.control.fusion-2010,
willis-all-cop-phys.rev.e-2011, hutchinson-phys.rev.e-2012}.
There have also been measurements of the interaction between dust particles~\cite{konopka-phd, 
konopka-mor-rat-phys.rev.lett-2000, hebner-ril-mar-phys.rev.e-2003},
but it is difficult to judge on the accuracy of wake models
because of experimental uncertainties, limited measurement range of distances, and poorly 
known parameters in the levitation region (see, e.g., Figs.~1 and 2 of Ref.~\cite{kompaneets-kon-ivl-phys.plasmas-2007} and 
Fig.~3 of Ref.~\cite{kompaneets-vla-ivl-new.j.phys-2008}, where the same measurements were fitted by quite different models by 
adjusting model parameters).

While theoretical and simulation efforts to 
describe the wakes generated by dust particles
have been focused 
on the homogeneous plasma approximation,
the levitation region is usually considerably inhomogeneous,
as evidenced by measurements of the resonance 
frequency of vertical oscillations~\cite{kompaneets-kon-ivl-phys.plasmas-2007, konopka-phd}. 
The measured frequency $f_{\rm res}=\omega_{\rm res}/(2\pi)$ yields
the field inhomogeneity length $\simeq g/\omega_{\rm res}^2$ 
(with the ion drag force~\cite{morfill-ivl-rev.mod.phys-2009} and dust charge variations~\cite{morfill-ivl-rev.mod.phys-2009} being neglected), 
which turns out to be about 
the characteristic shielding length in the region
(see, e.g., Sec. VB of Ref.~\cite{kompaneets-kon-ivl-phys.plasmas-2007}).

To the best of our knowledge, there have been no studies of the effect of the inhomogeneity on wake properties.
Presumably, this is because the inhomogeneity is challenging to account for: The standard 
calculation method based on
the three-dimensional Fourier transform in space and the dielectric function becomes inapplicable,
and the resulting equations entail substantial numerical difficulties.

In this paper, we address the role of the inhomogeneity by rigorously calculating the point-charge potential
in the collisionless Bohm sheath~\cite{riemann-j.phys.d.appl.phys-1991} 
(which is one of the best-known ``simple'' models for an inhomogeneous plasma with ion flow)
and comparing the results with the homogeneous approximation. Here, ``rigorously'' means
that we calculate the exact potential, making {\it no further approximation} in addition
to the common linear perturbation approximation. As the collisionless Bohm sheath is a model that has a number of
well-known limitations (see Sec.~\ref{discussion}),
our study is not intended to
precisely describe how dust particles are shielded under typical experimental conditions.
Our results, however, indicate the essential qualitative changes introduced by the inhomogeneity, which we believe 
to be the generic features characterizing wakes in inhomogeneous plasma flows.

We hope that this paper will be also interesting in that we 
describe a working method to accurately and effectively 
calculate the potential due to an extraneous charge in an inhomogeneous plasma 
(see Appendix~\ref{technical-details}).
Implementation of the method is relatively simple, so it may be further utilized for various inhomogeneous plasma environments.

\section{Model}

\subsection{Basic equations}

\label{basic-equations}

\begin{figure}
\includegraphics[width=8 cm]{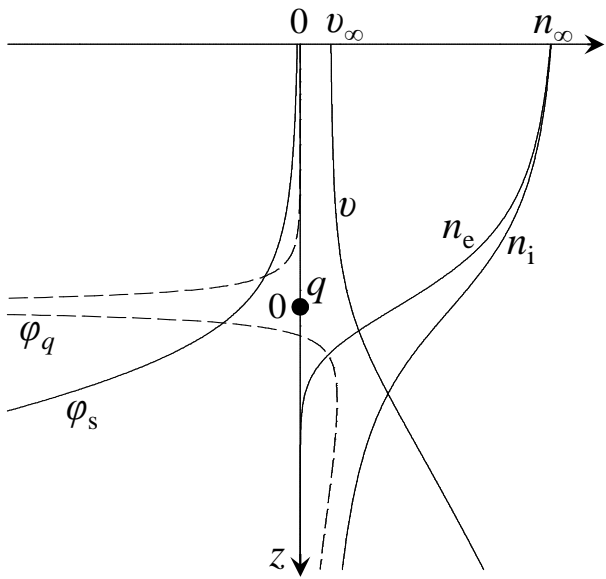}
\caption{Sketch of the problem. The solid curves illustrate the unperturbed sheath, showing the
electric potential $\varphi_{\rm s}(z)$, ion number density $n_{\rm i}(z)$, electron number density $n_{\rm e}(z)$, and ion flow velocity $v(z)$.
The dashed line shows the potential perturbation $\varphi_q$ (along the $z$ axis) due to the immersed charge $q<0$.}
\label{sketch}
\end{figure}

We consider a point non-absorbing charge $q$ located at ${\bf r}={\bf 0}$ and immersed in 
an inhomogeneous plasma consisting of Boltzmann electrons and cold flowing singly ionized ions.
A sketch of the problem
is shown in Fig.~\ref{sketch}.
We assume that at $z=-\infty$,
the plasma is homogeneous and has a number density $n_{\infty}$ and an ion flow velocity ${v}_{\infty}$ directed in the positive $z$-direction.
We set the electric potential at $z=-\infty$ equal to zero and assume that the 
unperturbed electric potential $\varphi_{\rm s}(z)$
takes a value $\varphi_0<0$ at $z=0$. 
(The subscript ``s'' stands for ``sheath''; ``unperturbed'' refers, here and in the following, to the state in the absence of the charge $q$.) 
No wall is included 
in our model,
as we assume that the wall towards which the unperturbed flow is directed~\cite{riemann-j.phys.d.appl.phys-1991}
is located sufficiently far from the charge; we adopt this assumption in order 
to investigate the pure effect of the inhomogeneity rather than 
the combined effect of the inhomogeneity and proximity of the wall.
Note that
in complex plasma experiments, the particles are usually negatively charged, but in our model the sign of $q$ is unimportant
in view of the linear perturbation approximation introduced below.

As illustrated in Fig.~\ref{sketch}, in the unperturbed sheath, 
(i) the ion flow velocity 
increases with $z$ as ions are accelerated by the electric field, 
(ii) the ion density decreases with $z$ to keep the ion flux constant, and (iii) the 
electron density decreases with $z$ faster than the ion density
to ensure a positive net charge density, resulting in the field being directed in the positive $z$-direction.

In the presence of the charge, the system in its steady state is described by the ion continuity equation
\begin{equation}
\label{continuity}
\nabla \cdot (n_{\rm i} {\bf v})=0,
\end{equation}
ion momentum equation
\begin{equation}
m\left( {\bf v} \cdot \nabla \right) {\bf v} = - e \nabla \varphi,
\end{equation}
Boltzmann distribution of electrons
\begin{equation}
\label{boltzmann}
n_{\rm e} = n_{\infty} \exp \left( -\frac{e \varphi}{T_{\rm e}} \right),
\end{equation}
and Poisson's equation
\begin{equation}
\label{poisson}
-\nabla^2 \varphi = 4\pi \left[ e (n_{\rm i} - n_{\rm e}) + q\delta({\bf r}) \right],
\end{equation}
where $n_{\rm i}$ and $n_{\rm e}$ are the ion and electron number densities, respectively,
${\bf v}$ is the ion flow velocity, 
\begin{equation}
\label{potential-definition}
\varphi=\varphi_{\rm s} +\varphi_q
\end{equation}
is the electric potential, with
$\varphi_q$ being the potential perturbation due to the charge (the wake potential),
$T_{\rm e}$ is the electron temperature, $m$ is the ion mass, $e$ is the elementary charge,
and $\delta({\bf r})$ is the delta-function. The unperturbed system is described by Eqs.~(\ref{continuity})-(\ref{poisson}) 
with $q=0$. We follow the common assumption \cite{riemann-j.phys.d.appl.phys-1991} that the velocity $v_{\infty}$ is the Bohm velocity, 
\begin{equation}
\label{initial-velocity}
v_{\infty}=\sqrt{\frac{T_{\rm e}}{m}}.
\end{equation}

Below we focus on the wake potential $\varphi_{q}({\bf r})$.
The problem is solved in the linear approximation (its applicability being discussed in Sec.~\ref{discussion}), i.e., 
Eqs.~(\ref{continuity})-(\ref{poisson}) are linearized with respect to the perturbations induced by the charge.

\subsection{Solution}
\label{solving-method}
We first make the following transformation to rewrite equations in a dimensionless form:
\begin{eqnarray}
\frac{\bf r}{\lambda_{\rm e \infty}} \to {\bf r}, \quad \frac{n_{\rm i}}{n_{\infty}} \to {n}_{\rm i}, \nonumber \\
\frac{\bf v}{v_\infty} \to {\bf v}, \quad
\psi=-\frac{e\varphi}{T_{\rm e}}, 
\end{eqnarray}
where
\begin{equation}
\lambda_{{\rm e} \infty} 
=\sqrt{      \frac{ T_{\rm e}}{4\pi n_{\infty}e^2}   }
\end{equation}
is the electron Debye length at $z=-\infty$
and $v_\infty$ is given by Eq.~(\ref{initial-velocity}). 
In these dimensionless notations, Eqs.~(\ref{continuity})-(\ref{poisson}) take the form
\begin{equation}
\label{continuity-normalized}
\nabla \cdot ({n}_{\rm i} {\bf v})=0,
\end{equation}
\begin{equation}
\label{momentum-normalized}
\left( {\bf v} \cdot \nabla \right) {\bf v} =\nabla \psi, \\
\end{equation}
\begin{equation}
\label{poisson-normalized}
\nabla^2 \psi = {n}_{\rm i} - \exp(-\psi) + A\delta({\bf r}),
\end{equation}
where 
$A= {q}/(en_{\infty}\lambda^3_{{\rm e} \infty})$.
It is convenient to introduce the dimensionless control parameter
\begin{equation}
\psi_0=-\left. \frac{e\varphi_{\rm s}}{ T_{\rm e}} \right|_{z=0},
\end{equation}
characterizing the charge location relative to the unperturbed plasma structure.

Calculating the unperturbed variables is nothing but solving the collisionless Bohm 
sheath model~\cite{riemann-j.phys.d.appl.phys-1991}.
The unperturbed momentum and continuity equations for ions
yield
\begin{equation}
\label{velocity-density-unperturbed}
{v}_{}=\sqrt{1+2\psi_{}}, \quad {n}_{\rm i}=\frac{1}{\sqrt{1+2\psi_{}}}.
\end{equation}
We substitute Eq.~(\ref{velocity-density-unperturbed}) into the unperturbed Poisson's equation (\ref{poisson-normalized}), 
multiply it by $d\psi_{}/d{z}$,
and integrate the resulting equation over ${z}$ from ${z}=-\infty$
to an arbitrary ${z}$ using the boundary conditions 
$\left. 
\psi_{} 
\right|_{ z \to -\infty} = 0$ and $\left.d \psi /dz \right|_{ z \to -\infty} = 0$.
This yields a first-order differential equation for $\psi_{}( z)$, whose solution for the boundary condition
$\left. \psi_{} \right|_{ z=0}=\psi_0$
is given by
\begin{equation}
\label{potential-unperturbed}
 z=\frac{1}{\sqrt{2}}\int_{\psi_0}^{\psi_{}} \frac{d\psi_{}}{  \sqrt{ \sqrt{1+2\psi_{}} + \exp(-\psi_{}) -2 }}.
\end{equation}
From this equation we find $\psi_{}( z)$ numerically, which is further used to calculate ${v}_{}({z})$ and ${n}_{\rm i}({z})$
from
Eq.~(\ref{velocity-density-unperturbed}). Note that 
the solid lines illustrating the unperturbed sheath in Fig.~\ref{sketch} are obtained by the exact calculation 
for $\psi_0=1$ (the range $-10 < z < 10$ is shown).
Choosing a different value of $\psi_0$
merely results in a shift of these curves along the $z$ axis. 

To calculate the perturbations, we use the two-dimensional Fourier transform with respect to ${\bf r}_\perp$, which is the component of ${\bf r}$ 
perpendicular to the $z$ axis. 
We linearize Eqs.~(\ref{continuity-normalized})-(\ref{poisson-normalized}) with respect to the perturbations
and take the Fourier transform of the resulting equations [i.e., we
multiply them by $\exp(-i {\bf k}_\perp \cdot {\bf r}_\perp)$, where ${\bf k}_\perp$ is a vector perpendicular to the $z$ axis,
and integrate them over ${\bf r}_\perp$]. We arrive at
\begin{equation}
\label{continuity-perturbations}
\frac{d}{d{z}} \left( \hat{{n}}_{\rm i} {v}_{} 
+{n}_{\rm i} \hat{{v}}_{z}\right)
+i {k}_\perp {n}_{\rm i} \hat{{v}}_{x}=0,
\end{equation}
\begin{equation}
\label{momentum-perturbations-z}
\frac{d (v \hat{v}_z)}{d z} =\frac{d \hat{\psi}}{d z},
\end{equation}
\begin{equation}
\label{momentum-perturbations-x}
v \frac{d \hat{v}_x}{d z}= i {k}_\perp \hat{\psi},
\end{equation}
\begin{equation}
\label{poisson-perturbations}
-k_\perp^2 \hat{\psi}+\frac{d^2 \hat{\psi}}{d z^2}=\hat{n}_{\rm i}+\exp(-\psi)\hat{\psi}+A\delta(z),
\end{equation}
where $\hat{{n}}_{\rm i}$, $\hat{v}_z$, $\hat{{v}}_{x}$, and $\hat{\psi}$ are the Fourier-transformed perturbations,
while ${n}_{\rm i}$, $v$, and $\psi$ are the unperturbed quantities
given by Eqs.~(\ref{velocity-density-unperturbed}) and (\ref{potential-unperturbed}).


Our next steps, described in detail in Appendix~\ref{technical-details}, are 
(i) to reduce
Eqs.~(\ref{continuity-perturbations})-(\ref{poisson-perturbations}) to a single equation for $\hat \psi$,
(ii) to rewrite this equation using the normalization
\begin{equation}
\label{phi_q-normalization}
\frac{\varphi_q}{q/\lambda_{\rm e \infty}} \to \varphi_q,
\end{equation}
which cancels out the parameter $A$,
(iii) to determine the physically correct boundary condition
by taking into consideration the  Landau damping,
(iv) to numerically solve the equation using the above boundary condition, and
(v) to numerically inverse Fourier-transform the result.

\subsection{Reference point: Wake in a homogeneous plasma}
\label{reference-point}
To see the effect of the plasma inhomogeneity on $\varphi_{q}({\bf r})$, 
we compare our results with those obtained in the homogeneous model in which
the unperturbed plasma is assumed to have the same electron Debye length, ion density,
and ion flow velocity as those in the sheath at $z=0$. The potentials $\varphi_q({\bf r})$ derived from the two models
must coincide as $\psi_0 \to 0$, which is one of the tests we used to ensure the correctness of our calculations.

In the homogeneous model, the unperturbed potential $\varphi_{\rm s}$ is obviously constant
and the potential perturbation is~\cite{montgomery-joy-sug-plasma.phys-1968}
\begin{equation}
\label{potential-homogeneous}
\varphi_{q}({\bf r})=\frac{q}{r}+ \frac{q}{2\pi^2} \int d{\bf k} \, \frac{\exp(i {\bf k}\cdot {\bf r})}{k^2} \left( \frac{1}{D({\bf k})} - 1\right),
\end{equation}
where
\begin{equation}
\label{dielectric-function}
D({\bf k})= 1+\frac{1}{(\lambda_{{\rm e} 0}k)^2}-\frac{\omega_{\rm pi 0}^2}{({\bf k}\cdot{\bf v}_0-i0^{+})^2}
\end{equation}
is the static dielectric function.
The relevant electron Debye length and ion plasma frequency are
\begin{equation}
\label{electron-Debye}
\lambda_{{\rm e} 0}=\sqrt{      \frac{ T_{\rm e}}{4\pi n_{{\rm e} 0}e^2}   }, \quad 
\omega_{\rm pi 0}=\sqrt{\frac{4\pi n_{\rm i 0}e^2}{m}},
\end{equation}
respectively, and the subscript ``0'' denotes the unperturbed quantities taken from our inhomogeneous
model at $z=0$.
The term $-i0^{+}$ (where $0^{+}$ is an infinitesimal positive number) 
represents an infinitely small Landau damping~\cite{kompaneets-vla-ivl-new.j.phys-2008},
which is important as it removes the singularity of the integrand in Eq.~(\ref{potential-homogeneous}), 
with the minus sign
in $-i0^{+}$ resulting in the downstream location of the oscillatory wake structure \cite{vladimirov-ish-phys.plasmas-1996}.
This makes it obvious that in our inhomogeneous model, we do need 
to take into consideration the Landau damping, 
which is step (iii) mentioned in Sec.~\ref{solving-method}.
The calculation of the integral~(\ref{potential-homogeneous}) is detailed in Appendix~\ref{calculation-homogeneous}.

\section{Results}
\label{results}
\begin{figure}
\includegraphics[width=8 cm]{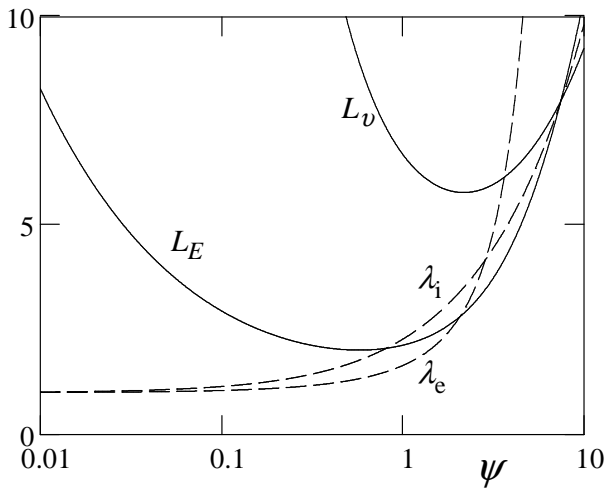}
\caption{The velocity and field inhomogeneity 
lengths, $L_v$ and $L_E$ [Eq.~(\ref{inhomogeneity-lengths})], as well as the electron Debye length $\lambda_{\rm e}$
and ion shielding length $\lambda_{\rm i}$ [Eq.~(\ref{ion-shielding-length})]. The lengths are normalized by $\lambda_{{\rm e} \infty}$
and shown as functions of the normalized sheath potential $\psi$.}
\label{inhomogeneity}
\end{figure}

One may expect the shielding cloud 
to be considerably affected by the inhomogeneity
when the respective spatial scales are comparable.
Therefore, before providing our results, we show in Fig.~\ref{inhomogeneity} the following 
unperturbed quantities: 
the electron Debye length and ion Debye length defined using the ion kinetic energy~\cite{kompaneets-vla-ivl-new.j.phys-2008},
\begin{equation}
\label{ion-shielding-length}
\lambda_{\rm e}=\sqrt{      \frac{ T_{\rm e}}{4\pi n_{\rm e}e^2}   },\quad
\lambda_{\rm i}=\sqrt{\frac{ mv^2}{4\pi n_{\rm i}e^2}},
\end{equation}
as well as the velocity and field inhomogeneity lengths,
\begin{equation}
\label{inhomogeneity-lengths}
L_v=v\left( \frac{d v}{dz} \right)^{-1}, \quad
L_E=E\left( \frac{d E}{dz} \right)^{-1},
\end{equation}
where $E=|d\varphi_{\rm s}/dz|$ is the sheath electric field. 
We see that the inhomogeneity is weak at small $\psi$ and becomes substantial at $\psi \simeq 1$---$3$, depending on
whether $L_v$ or $L_E$ is considered.

\begin{figure}
\includegraphics[width=8 cm]{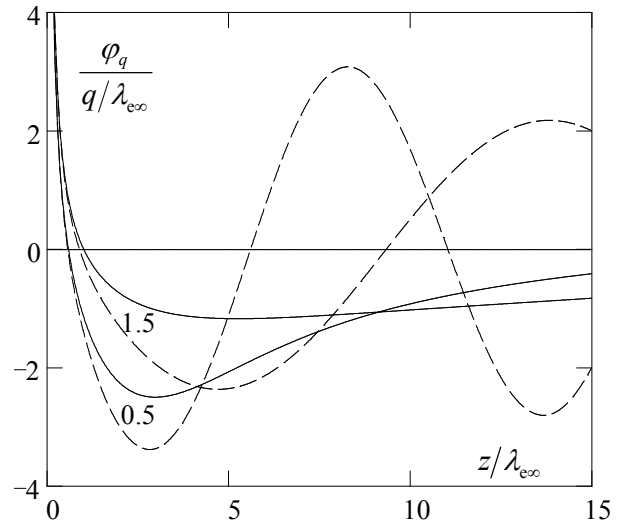}
\caption{Potential perturbation in the downstream direction, $\varphi_q(r_\perp=0,z)$. 
The solid and dashed lines represent the results of the inhomogeneous and 
homogeneous models, respectively. The numbers near the curves indicate the values of $\psi_0$.
The graph is obtained by truncating the integration at $k_\perp=20$ (see text).}
\label{along}
\end{figure}

Let us start with the potential perturbation in the downstream direction 
(i.e., for $z>0$ and $r_\perp=0$).
For this direction,
the integral over $k_\perp$ determining $\varphi_q$ [see Eqs.~(\ref{inverse-transform-final}) 
and (\ref{homogeneous-final})]
logarithmically diverges at large $k_\perp$ in both the inhomogeneous and homogeneous models,
the divergence being due to the cold ion 
approximation~\cite{kompaneets-vla-ivl-new.j.phys-2008}
and disappearing for $r_\perp \not = 0$ or $z<0$. 
For $z>0$ and $r_\perp=0$, we truncate the integration at a certain large number, $k_\perp=20$,
and show the results in Fig.~\ref{along}.
We see that the oscillatory wake structure, always present
in the homogeneous model, entirely disappears in the inhomogeneous model when $\psi_0$ is still considerably
less than unity.
The minimum in the potential perturbation is considerably more shallow in the inhomogeneous model.
Interestingly, the minimum location is practically unaffected by the inhomogeneity.
In the limit $\psi_0 \to 0$, our numerical calculations for both models
yield exactly matching oscillatory structures.

\begin{figure}
\includegraphics[width=8 cm]{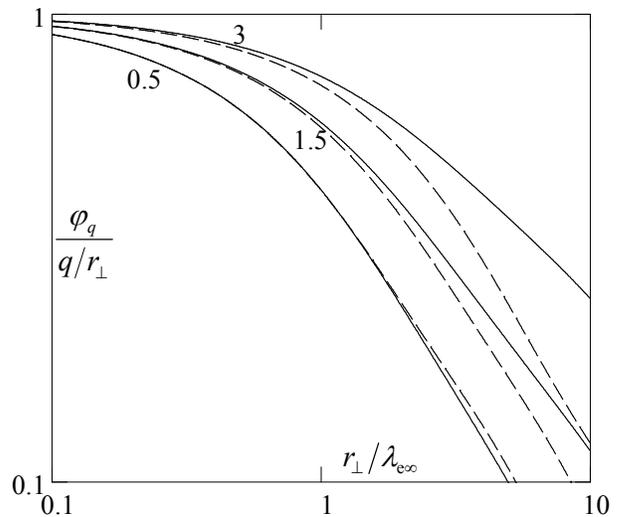}
\caption{Potential 
perturbation in the direction perpendicular to the flow, $\varphi_q(r_\perp,z=0)$, divided by the Coulomb potential
and shown on a log-log graph.
The line styles and numbers near the curves
bear the same meanings as in Fig.~\ref{along}.}
\label{perpendicular}
\end{figure}

Figure~\ref{perpendicular} shows the potential perturbation in the direction perpendicular to the flow. It is seen 
to be repulsive (for a charge of the same sign)
and not dramatically affected by the inhomogeneity up to
$\psi_0 \sim 3$, starting from which the inhomogeneity results in a substantially weaker screening.

\begin{figure}
\includegraphics[width=8 cm]{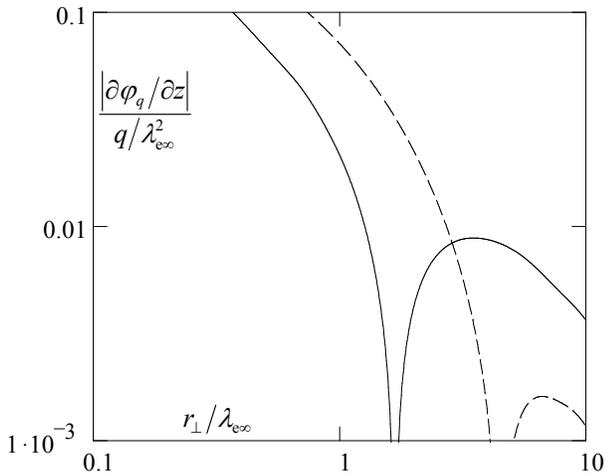}
\caption{The derivative $\left. (\partial \varphi_q/\partial z) \right|_{z=0}$ as a function of $r_\perp$. Shown are the results for $\psi_0 = 1.5$, the line styles
bear the same meanings as in Fig.~\ref{along}.}
\label{dphidz}
\end{figure}

Figure~\ref{dphidz} shows the absolute value of the derivative $\left. (\partial \varphi_q/\partial z) \right|_{z=0}$ as a function of $r_\perp$. 
This quantity, as noted in Sec.~\ref{discussion}, is of interest
in the context of the mode coupling instability~\cite{ivlev-mor-phys.rev.e-2001, couedel-nos-ivl-phys.rev.lett-2010, couedel-zhd-ivl-phys.plasmas-2011,
rocker-ivl-kom-phys.plasmas-2012} 
observed in two-dimensional plasma crystals. 
Here we see a remarkably strong effect of the inhomogeneity 
at $\psi_0 = 1.5$: The magnitude of the derivative is strongly reduced
at $r_{\perp}\sim \lambda_{\rm e \infty}$ (corresponding to a typical interparticle distance),
and the sign changes at a much smaller $r_\perp$ than in the homogeneous case.

\begin{figure*}
\includegraphics[width=15 cm]{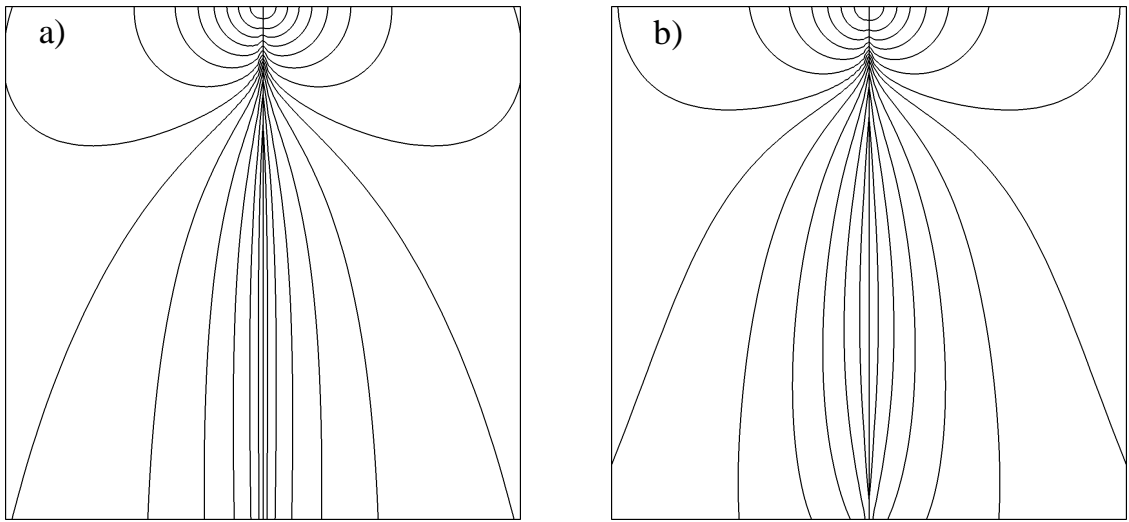}
\caption{Potential perturbation $\varphi_q(r_\perp, z)$ for (a) the inhomogeneous and (b) homogeneous models.
Shown are the equipotential lines for $\psi_0 = 1.5$,
the potential step is not kept constant.
The charge is located in the center of the upper edge,
the flow is directed downwards. The dimensions are $8 \lambda_{\rm e \infty} \times 8 \lambda_{\rm e \infty}$
in both cases.
In the approximation considered, in both cases the potential perturbation
becomes logarithmically infinite
at the line $r_\perp=0$ and $z>0$, which is also drawn.}
\label{5-4}
\end{figure*}

Figure~\ref{5-4} shows the contour plots of $\varphi_q(r_\perp, z)$ for 
(a) the inhomogeneous and (b) homogeneous models.
The plots are similar for the region 
shown, 
although not exactly coinciding (compare, e.g., the angles at which the equipotential 
lines cross the plane $z=0$ as well as the behavior of the equipotential lines at large $z$ near the $z$ axis).

\section{Discussion and conclusions}
\label{discussion}

\subsection{Model assumptions}
To draw conclusions, let us first discuss the relevance of 
our model to experiments with complex plasmas.

(i) The sheath is meant to be only a part of the plasma-wall transition layer separating 
an isotropic plasma from an electrode~\cite{riemann-j.phys.d.appl.phys-1991}.
It is often unclear whether the dust particles in a given experiment
are levitated in the sheath, presheath, or at the boundary between the two.
An estimate of the electron-to-ion density ratio
at the levitation position for a typical experiment~\cite{konopka-phd, 
konopka-mor-rat-phys.rev.lett-2000}, 
based on Poisson's equation and the measured resonance frequency of vertical particle oscillations,
yields $\simeq 0.85$~\cite{kompaneets-kon-ivl-phys.plasmas-2007}, suggesting that particles were levitated near the boundary.
Sufficiently heavy particles, especially under hypergravity
conditions~\cite{beckers-ock-wol-phys.rev.lett-2011}, may be levitated in the sheath.
It is noteworthy that while the Bohm sheath model represents a dc regime, most dusty plasma experiments are performed
in rf discharges, where electrons respond to the rf field. 
It is thus clear that the Bohm sheath
is not a precise model to describe the levitation region
for most experiments. However, to qualitatively appreciate the effect of the inhomogeneity
on the wake, we only need a self-consistent plasma profile that roughly resembles the actual one,
and the Bohm sheath model is certainly adequate for such a purpose.

(ii) Ions in the levitation region are generally not cold, as they experience collisions on their way through the presheath, 
forming a velocity distribution with a suprathermal width~\cite{zeuner-mei-vacuum-1995, kompaneets-ivl-vla-phys.rev.e-2012, 
lampe-roc-joy-phys.plasmas-2012, kompaneets-tys-vla-phys.plasmas-2013}. The latter should
affect the shielding, as discussed in Sec.~\ref{effects-of-the-inhomogeneity} in the context of the results of this paper ---
but, again, to specifically identify the role of the inhomogeneity,
it seems reasonable to start with the assumption of cold ions. Moreover, 
measurements of the ion
velocity distribution at the electrode
show that depending on the pressure and rf power, the characteristic width of the distribution can 
be considerably smaller than the flow velocity~\cite{zeuner-mei-vacuum-1995}, making the cold ion approximation 
quite reasonable for a certain range of distances from the electrode.

(iii) A necessary condition for the model of the collisionless Bohm sheath to be applicable is that
the ion-neutral collision length must be much larger than the electron Debye length~\cite{riemann-j.phys.d.appl.phys-2003}. 
For a typical experiment performed at 2.7 Pa~\cite{konopka-phd, 
konopka-mor-rat-phys.rev.lett-2000}, their ratio is $\simeq 5$~\cite{kompaneets-kon-ivl-phys.plasmas-2007}.
Hence, the collisionless approximation should be quite reasonable for  
experiments performed at lower pressures~\cite{nosenko-ivl-zhd-phys.plasmas-2009, couedel-nos-zhd-phys.rev.lett-2009}.

(iv) We use the linear perturbation approximation. Nonlinear effects can indeed be substantial
for experiments in which dust particles are levitated in a more or less isotropic region.
However, for ion flow velocities of the order of the Bohm velocity,
nonlinear effects should be insignificant as the Coulomb radii for ions and electrons, 
defined as
\begin{equation}
\label{coulomb-radii}
R_{\rm i} = \frac{|q|e}{mv^2}, \quad R_{\rm e} = \frac{|q|e}{T_{\rm e}},
\end{equation}
are usually much smaller than the Debye lengths $\lambda_{\rm i,e}$ [defined by Eq.~(\ref{ion-shielding-length})]. 
Nevertheless, accounting for the presence 
of low-energy ions due to ion-neutral collisions
may still affect the results to some extent~\cite{hutchinson-haa-phys.plasmas-2013}.

(v) The assumption of Boltzmann electrons becomes invalid near the wall because of the absorption.
However, our model is fully self-consistent as we assume the wall to be located sufficiently far from the charge.
In experiments with complex plasmas in rf discharges, the levitation height above the electrode
is usually quite large, e.g., an order of magnitude larger than the electron Debye length~\cite{konopka-phd}.

(vi) Finally, we assume a non-absorbing point charge. The absorption should be 
generally insignificant for flow velocities of the order of the Bohm velocity
because in this regime
$\lambda_{\rm i,e}$ are typically two orders of magnitude larger than
the ion and electron collection radii (impact parameters)~\cite{morfill-ivl-rev.mod.phys-2009}. The assumption of a point charge is justified
for the same reason --- the dust size is typically two orders of magnitude smaller than $\lambda_{\rm i,e}$.

Now that the relevance of our model has been discussed, let us elaborate on our findings.

\subsection{Effects of the inhomogeneity}
\label{effects-of-the-inhomogeneity}
The first effect is the disappearance of the oscillatory wake structure.
We note that the presence of these oscillations within the assumption of a homogeneous plasma
is model-dependent.
For instance, the number of oscillations
is infinite for cold ions and Boltzmann electrons
but becomes finite when the ion distribution is a shifted Maxwellian 
(in which case the far-field potential exhibits a monotonic $r^{-3}$-decay~\cite{montgomery-joy-sug-plasma.phys-1968}).
In the latter model, the number of the oscillations
is determined by the flow-to-thermal velocity ratio as well as the ratio of
the temperatures.
When the first ratio does not exceed a certain value of the order of unity, the wake exhibits a single potential well,
at least when the electron-to-ion temperature ratio is infinitely large~\cite{peter-j.plasma.phys-1990}.
Hence, a homogeneous model with a realistic ion velocity distribution
(whose characteristic width is comparable
to the ion flow velocity~\cite{zeuner-mei-vacuum-1995,kompaneets-ivl-vla-phys.rev.e-2012})
may also yield a non-oscillatory wake.
The issue is complicated by a non-Maxwellian form of the ion velocity distribution,
ion-neutral collisions, and the electric field.
A model accounting for all these factors can still yield an oscillatory wake,
as follows from Eq. (6) of Ref.~\cite{kompaneets-kon-ivl-phys.plasmas-2007}.

The present study suggests that the inhomogeneity tends to suppress the oscillations. 
Since the inhomogeneity
is usually significant in experiments (as noted in Sec.~\ref{introduction} and discussed in Sec.~\ref{implications}),
one can expect the wake oscillations unlikely to be formed in most experiments,
regardless of what homogeneous models predict.

To shed a light on how the oscillations are suppressed by the inhomogeneity, let us
focus on the large-$z$ behavior of the
Fourier-transformed (over ${\bf r}_\perp$) potential perturbation, considering the case $k_\perp=0$ for simplicity. 
This behavior is described by
\begin{equation}
\label{oscillations}
\frac{d^2 \hat \varphi_q}{dz^2}+\frac{n_{\rm i}}{v^2} \hat \varphi_q =0,
\end{equation}
as follows from Eq.~(\ref{final-equation}).
This has a form of the oscillator equation with a variable frequency. The latter
is determined by the unperturbed ion profile [Eqs.~(\ref{velocity-density-unperturbed}) and (\ref{potential-unperturbed})],
while the electrons do not contribute, as the Boltzmann factor $\exp(-\psi)$
becomes exponentially small at large $z$.
Obviously, if $n_{\rm i}/v^2$ is considered to be constant, Eq.~(\ref{oscillations}) yields an oscillatory structure
of $\hat \varphi_q$.
For the collisionless Bohm sheath, at large $z$ we have
$n_{\rm i}/v^2 = (2/9)z^{-2}$ (plus the higher-order terms),
as follows from Eqs.~(\ref{velocity-density-unperturbed}) and (\ref{potential-unperturbed}).
For this dependence, the general solution of Eq.~(\ref{oscillations}) becomes non-oscillatory,
$\hat \varphi_q = C_1 z^{1/3} +C_2 z^{2/3}$, indicating
that the inhomogeneity tends to inhibit wake oscillations.

The second effect is that the wake becomes considerably weaker, 
i.e., $\varphi_q$ dips to a less negative value (see Fig.~\ref{along}).
This effect is not obvious.
On the one hand, in comparison to the homogeneous case,
the plasma at $z>0$ (where the wake is formed) has smaller 
ion and electron densities and a larger flow velocity
and thus may be considered as less capable of considerable ``overshielding'', so one could indeed expect a weaker wake. 
But on the other hand, the ion deflection
starts long before 
the ions reach the plane $z=0$. At this level they already have a transverse velocity,
which could be larger in the inhomogeneous model (because at $z<0$
the ions are slower and thus more prone to the deflection than in the homogeneous case),
resulting in a larger transverse displacement and hence a stronger wake. 
Our results imply that the former
contribution is stronger than the latter one.

Yet another effect is a substantially weaker screening of the Coulomb potential in the perpendicular direction (for a sufficiently 
strong inhomogeneity). Note that for very large $\psi_0$,
the homogeneous model predicts a long-range attraction (for a charge of the same sign) in this direction~\cite{kompaneets-vla-ivl-new.j.phys-2008}.
In our inhomogeneous model the attraction has not been found,
although we did not study the regime of unrealistically large $\psi_0$ ($\gg 3$).
Also note that such an attraction
was obtained, for a certain set of parameter values,
in the study of Ref.~\cite{ignatov-plasma.phys.rep-2004} including the inhomogeneity
along with ion-neutral collisions, ionization, and close proximity of the wall. Since we found the inhomogeneity to weaken the
screening in the perpendicular direction, we suggest that it generally
weakens (or eliminates, depending on conditions) the attraction in this direction.

On the other hand, some wake parameters are practically unaffected by the inhomogeneity, for instance,
the location of the wake focus
(i.e., of the minimum of the wake potential).

The same applies to the ion drag force
${\bf F}_{\rm dr}=-q \left. \nabla \varphi_{q} \right|_{{\bf r}={\bf 0}}$ (which is the electric force exerted by 
the point charge $q$ on itself through the plasma perturbation).
Indeed, the ion drag force can be written in the dimensionless variables by using Eq.~(\ref{inverse-transform-final}) as
\begin{equation}
\label{ion-drag-integral}
F_{{\rm dr}}= \int_0^{R_{\rm i}^{-1}} dk_\perp \, 
k_\perp \left( \frac{1}{2\pi}\left. \frac{d \hat \varphi_q}{dz} \right|_{z=0^-} -1 \right),
\end{equation}
where we avoid the well-known logarithmic divergence of the ion drag force~\cite{peter-j.plasma.phys-1990} 
by classically truncating the integration 
at $k_\perp=R_{\rm i}^{-1}$ (where
nonlinear effects become significant).
Here, $R_{\rm i}$ is defined by Eq.~(\ref{coulomb-radii}) and normalized by $\lambda_{\rm e \infty}$,
the ion drag force $F_{{\rm dr}}$ is normalized by $q^2/\lambda_{\rm e \infty}^2$,
and $0^{-}$ is an infinitesimal negative number.
Since the normalized $R_{\rm i}^{-1}$ is usually very large (which is the condition to employ
the linear perturbation approximation),
the ion drag force is primarily determined by the coefficient
of the $k_\perp^{-1}$-dependence of the integrand in Eq.~(\ref{ion-drag-integral}) (at large $k_\perp$). We have found this coefficient to
be {\it exactly} the same for the inhomogeneous and homogeneous models, which reflects
the obvious fact that the plasma inhomogeneity is negligible at small spatial scales ($\sim R_{\rm i}$).

\subsection{Implications}
\label{implications}

A natural question arises as to what extent the approximation of a homogeneous plasma 
is accurate to describe wakes
in experiments with complex plasmas. Obviously, the levitation position and hence the local magnitude of the inhomogeneity depend
on the particle size, so let us make some estimates. 

We use the collisionless Bohm sheath model in conjunction with
the vertical force balance $-qE=Mg$ (where $M$ is the particle mass),
neglecting the ion drag
for the moment. This results in the following expression
for the particle radius $a(\psi_0)$ as a function of the sheath potential at the levitation height:
\begin{equation}
\label{vertical-balance}
a^2(\psi_0)=\frac{3 T_{\rm e}^2 z_{\rm d}(\psi_0)}{2 \sqrt{2} \pi \rho g e^2\lambda_{\rm e \infty}}\sqrt{ \sqrt{1+2\psi_{0}} + \exp(-\psi_{0}) -2 },
\end{equation}
where $\rho$ is the particle material mass density
and $z_{\rm d}(\psi_0)=-q e/(aT_{\rm e})$ is the normalized dust charge, 
which we find from the charging equation $I_{\rm i}=I_{\rm e}$.
We assume the ion and electron fluxes on the particle, $I_{\rm i,e}(\psi_0)$, to be given by the orbit-motion-limited (OML) theory
for cold flowing ions and Maxwellian electrons~\cite{fortov-ivl-khr-phys.rep-2005,
morfill-ivl-rev.mod.phys-2009}, which yields
\begin{equation}
\label{charging}
\left(1+\frac{2z_{\rm d}(\psi_0)}{1+2\psi_0}\right)\exp\left[z_{\rm d}(\psi_0)+\psi_0 \right]=\sqrt{\frac{8m}{\pi m_{\rm e}}},
\end{equation}
where $m_{\rm e}$ is the electron mass. Here, we used Eqs.~(\ref{boltzmann}) and (\ref{velocity-density-unperturbed}) to express 
the ion and electron densities as well as the flow velocity 
as functions of $\psi_0$. 
Additionally, we need to take into account the stability condition, 
\begin{equation} 
\label{stability}
\left. \frac{d}{dz} \left( z_{\rm d}[\psi(z)] \frac{d\psi(z)}{dz} \right) \right|_{z=0}>0.
\end{equation}
By analyzing Eqs.~(\ref{vertical-balance})-(\ref{stability}) for typical values $\rho=1.5$~g/cm$^3$, 
$\lambda_{\rm e \infty}=0.5$~mm, $T_{\rm e}=2$~eV, and an argon plasma~\cite{konopka-phd, 
konopka-mor-rat-phys.rev.lett-2000},
we get the largest dust radius for which the levitation is possible 
to be $a \simeq 6.6$~$\mu$m, which corresponds to $\psi_0\simeq 2.7$. 
For smaller particles the value of $\psi_0$ is lower, reaching
$0.5$ at $a \simeq 3.4$~$\mu$m, which is a typical dust radius for experiments.
At this point the inhomogeneity still affects the wake considerably (see Fig.~\ref{along}), so the dust size must be substantially 
further reduced for the homogeneous approximation to become accurate. Note that much smaller (sub-micron) particles should rather 
be levitated in the presheath,
where Eqs.~(\ref{vertical-balance}) and (\ref{charging}) are inappropriate. 

To estimate the role of the ion drag,
we use the following expression:
\begin{equation}
\label{ion-drag-expression}
F_{\rm dr}=\frac{q^2\omega_{\rm pi 0}^2}{v_0^2}\ln \Lambda_0,
\end{equation}
where the pre-logarithmic factor is the coefficient of the $k_\perp^{-1}$-dependence 
of the integrand in Eq.~(\ref{ion-drag-integral}) at large $k_\perp$ (written in the dimensional variables), 
the Coulomb logarithm is 
taken to be 
\begin{equation}
\ln \Lambda_0=\ln \left( \frac{     {\rm min}(\lambda_{{\rm i} 0}, \, \lambda_{\rm e 0})  } {R_{\rm i 0}} \right),
\end{equation}
and the subscript ``0'' is used to explicitly refer to the particle location. By using Eq.~(\ref{ion-drag-expression}), we find the ratio 
of the ion drag to
gravity forces to be $\simeq 0.2$
at $\psi_0=0.5$. Since this ratio decreases with $\psi_0$, the ion drag should not considerably affect the vertical force balance unless
the particle is rather small.

These simple estimates indicate that for experiments with complex plasmas, 
homogeneous wake models are only accurate under rather special conditions,
e.g., if the particles are rather small or levitated by 
the thermophoresis force~\cite{rothermel-hag-mor-phys.rev.lett-2002}, gas flow~\cite{fink-zhd-tho-phys.rev.e-2012},
or under microgravity conditions~\cite{ivlev-mor-tho-phys.rev.lett-2008, 
khrapak-klu-hub-phys.rev.e-2012,
beckers-tri-kro-phys.rev.e-2013, khrapak-tho-cha-phys.rev.e-2013}. 
Speaking in terms of the inhomogeneity scale,
the velocity inhomogeneity length $L_v$ must be 
at least an order of magnitude larger
than $\lambda_{\rm i}$ (the length characterizing the collective ion response)
in order for the inhomogeneity to not affect the wake considerably (see Fig.~\ref{along} in 
conjunction with Fig.~\ref{inhomogeneity}).

By modifying the wakes, the plasma inhomogeneity affects a variety of static and dynamic dust structures.
For instance, the inhomogeneity effect on the derivative $\left. \partial\varphi_q/\partial z \right|_{z=0}$ (see Fig.~\ref{dphidz})
should influence the development of the mode-coupling instability in two-dimensional plasma crystals~\cite{ivlev-mor-phys.rev.e-2001, 
couedel-nos-ivl-phys.rev.lett-2010, couedel-zhd-ivl-phys.plasmas-2011, rocker-ivl-kom-phys.plasmas-2012}:
The growth rate of the instability is proportional to the square of the above derivative~\cite{kompaneets-ivl-tsy-phys.plasmas-2005},
so that the critical pressure (at which the instability is suppressed by the gas friction) dramatically
depends on the inhomogeneity. Furthermore, the weakening of the wake 
should affect the stability of vertical dust pairs~\cite{steinberg-sut-ivl-phys.rev.lett-2001, hebner-ril-mar-phys.rev.e-2003},
while the weakening of the screening in the 
perpendicular direction implies substantially stronger interparticle interactions in two-dimensional plasma crystals.

Yet another implication is that since the inhomogeneity tends to suppress the long-range attraction in the 
perpendicular direction (for a charge of the same sign),
it may be very difficult to experimentally realize a molecular-type interaction potential in two-dimensional plasma crystals.

\subsection{Conclusions}
We have demonstrated that 
the plasma inhomogeneity can dramatically modify the wake,
making it non-oscillatory and weaker.
We expect this to occur in many laboratory experiments with complex plasmas
as the inhomogeneity
in such experiments is usually quite significant.

\begin{acknowledgments}
The work was partially supported by the European Research Council
under the European Union's Seventh Framework Programme
(FP7 / 2007---2013) / ERC Grant agreement 267499.
\end{acknowledgments}

\appendix
\section{Calculation of $\varphi_q$ from Eqs.~(\ref{continuity-perturbations})-(\ref{poisson-perturbations})}
\label{technical-details}
In this Appendix, we explain steps (i)-(v) mentioned in Sec.~\ref{solving-method}.
Concerning step (i), we first express $\hat{v}_z$ and $\hat{v}_x$ via $\hat \psi$
by using Eqs.~(\ref{momentum-perturbations-z}) and (\ref{momentum-perturbations-x})
as well as the boundary conditions $\left. \hat{\bf v} \right|_{z=-\infty}={\bf 0}$ and $\left. \hat\psi\right|_{z=-\infty}=0$, which yields
\begin{equation}
\label{v_z}
\hat{v}_z = \frac{\hat \psi}{v} 
\end{equation}
and 
\begin{equation}
\label{v_x}
\hat{v}_x =ik_\perp \int_{-\infty}^{z} dz' \, \frac{\hat{\psi}}{v}.
\end{equation}
This allows us to express $\hat{n}_{\rm i}$ via $\hat \psi$, by substituting Eqs.~(\ref{v_z}) and (\ref{v_x})
into Eq.~(\ref{continuity-perturbations}) and integrating it from $z=-\infty$ to an arbitrary $z$ with
the boundary condition $\left. \hat{n}_{\rm i} \right|_{z=-\infty}=0$. By substituting the result into Eq.~(\ref{poisson-perturbations}), 
we obtain an integro-differential equation for $\hat \psi$. By rewriting it in terms of the normalized $\varphi_q$ [see Eq.~(\ref{phi_q-normalization})],
which is step (ii), we get
\begin{eqnarray}
\frac{\partial^2 \hat{\varphi}_q}{\partial z^2}=
\left( k_\perp^2 - \frac{n_{\rm i}}{v^2} +\exp(-\psi) \right)\hat{\varphi}_q \nonumber \\
+\frac{k_\perp^2}{v} \int_{-\infty}^z dz' \, n_{\rm i}(z') \int_{-\infty}^{z'} d{z''} \, \frac{\hat{\varphi}_q(z'')}{v(z'')} - 4\pi \delta(z),
\label{final-equation}
\end{eqnarray}
where $\hat \varphi_q$ is the Fourier-transformed (with respect to ${\bf r}_\perp$) potential perturbation
normalized as per Eq.~(\ref{phi_q-normalization}).

To find the physically correct boundary condition for Eq.~(\ref{final-equation}), which is step (iii), we employ the fact that
the physically correct solution $\hat \varphi_q(z)$ (when corrected for the Landau damping) must vanish at $z \to \pm \infty$.
At $z \to -\infty$,  Eq.~(\ref{final-equation}) becomes an equation with constant coefficients, 
so all its possible asymptotic solutions are linear combinations of $\exp(i k_{z*} z)$. The numbers 
$k_{z*}$ can be easily found 
analytically from Eq.~(\ref{final-equation}) to be two real roots
as well as two imaginary roots with opposite signs. In this limit,
the numbers $k_{z*}$ can also be obtained as the roots of the dielectric function (\ref{dielectric-function}),
where $n_{\rm e 0}$ and $n_{\rm i 0}$ are replaced by $n_{\infty}$, and $v_0$ by $v_\infty$:
\begin{equation}
\label{roots}
1+\frac{1}{k_{z*}^2+k_\perp^2}-\frac{1}{(k_{z*}-i0^{+})^2}=0.
\end{equation}
The term $-i0^{+}$ represents the Landau damping, as stated in Sec.~\ref{basic-equations}. By solving the above equation, we find that 
the Landau damping results in infinitesimal {\it positive} imaginary corrections to both ``real''  roots, meaning
that the corresponding solutions grow exponentially as $z \to -\infty$. Excluding these roots
as well as the imaginary root with a positive imaginary part, we get only one
$k_{z*}$ remaining. This $k_{z*}$ yields the following
long-distance behavior:
\begin{equation}
\label{asymptotic-behavior}
z \to -\infty: \quad \hat \varphi \propto \exp (\gamma z),
\end{equation}
where 
\begin{equation}
\gamma=       \sqrt{     \frac{k_\perp^2}{2} + k_\perp \sqrt{  1+ \frac{ k_\perp^2}{4}  }  }.  
\end{equation}

For the numerical integration, we 
convert Eq.~(\ref{final-equation}) into a system of first-order differential equations by introducing the following new variables:
\begin{eqnarray}
\label{notations-reducing}
\chi = \frac{d \hat \varphi_q}{dz}, \quad I_1=\int_{-\infty}^z dz' \, \frac{\hat \varphi_q(z')}{v(z')}, \nonumber \\
I_2=\int_{-\infty}^z dz' \, n_{\rm i}(z') I_1(z').
\end{eqnarray}
The resulting system (for $z \not = 0$) is
\begin{eqnarray}
\frac{d\chi}{dz}=\left( k_\perp^2 - \frac{n_{\rm i}}{v^2} +\exp(-\psi) \right)\hat{\varphi}_q 
+\frac{k_\perp^2}{v} I_2, \nonumber \\ 
\frac{d\hat \varphi_q}{dz}=\chi, \quad
\quad \frac{dI_1}{dz}=\frac{\hat \varphi_q}{v}, \quad \frac{dI_2}{dz}=n_{\rm i} I_1.
\label{system}
\end{eqnarray}
The delta-function in Eq.~(\ref{final-equation}) is replaced by the following condition:
\begin{equation}
\left. \chi \right|_{z=0^+}-\left. \chi \right|_{z=0^-}=-4\pi,
\end{equation}
while $\hat \varphi_q(z)$, $I_1(z)$, and $I_2(z)$ are continuous at $z=0$.

We set the starting point $z=z_-$ of the numerical integration of Eq.~(\ref{system})
to be a large negative number such that
varying the latter does not affect the potential in the region of interest.
(Similar variation tests are performed for all ``internal'' parameters
of the numerical procedure detailed below.)
We consider the value of $\hat \varphi_q(z_-)$ to be a certain number $\hat \varphi_{q-}$ (depending on $k_\perp$)
and use the following boundary conditions at $z=z_-$:
\begin{eqnarray}
\label{boundary-numerical}
z=z_-: \quad \chi= \gamma \hat \varphi_{q-}, \quad
I_1 = \hat \varphi_{q-} /\gamma, \quad I_2 = \hat \varphi_{q-} /\gamma^2,
\end{eqnarray}
which follow from Eqs.~(\ref{asymptotic-behavior}) and (\ref{notations-reducing}).
We find $\hat \varphi_{q-}$ by requiring that 
the numerical integration of the system~(\ref{system}) with the boundary conditions~(\ref{boundary-numerical})
yields $\hat \varphi_q=0$ at the end point $z=z_+$. The latter
is a large positive number such that varying it
does not affect the potential in the region of interest (similar to the starting point $z_-$).
Note that the condition $\hat \varphi_q(z_+)=0$ implies 
a conducting wall at $z=z_+$. We use the bisection method to find $\hat \varphi_{q-}$, initially choosing two 
arbitrary numbers $\hat \varphi_{q-}$ resulting in opposite signs of $\hat \varphi_q$ at $z=z_+$. 
For the integration of Eq.~(\ref{system}), we simply use Euler's method with a sufficiently small fixed integration step.

A difficulty arises at this point: 
It turns out that even a tiny relative difference between the ``upper'' and ``lower'' values of $\hat \varphi_{q-}$ obtained
by the above bisection method
results in a strong deviation between the corresponding ``upper'' and ``lower''
curves $\hat \varphi_q(z)$, which occurs already at quite small positive $z$ (unless $k_\perp$ is not small enough).
This does not prevent us from finding $\hat \varphi_{q-}$ rather precisely, but the problem is to achieve the 
sufficient accuracy
for $\hat \varphi_q(z)$.

To resolve the difficulty, we employ the following method. 
We stop the bisection procedure as soon as the difference between the ``upper'' and ``lower'' values 
of $\hat \varphi_{q-}$ becomes smaller than a certain threshold $\Delta_{\rm min}$,
and then integrate the system~(\ref{system}) to the point of the $z$ axis at which 
the deviation between the corresponding ``upper'' and ``lower'' curves $\hat \varphi_q(z)$
exceeds another threshold $\Delta_{\rm max}(>\Delta_{\rm min})$. 
The next step is to reduce the uncertainty of $\hat \varphi_q$ at this point
to $\Delta_{\rm min}$. 
For this, we first integrate Eq.~(\ref{system}) from this point using 
the middle values
of $\hat \varphi_q$, $\chi$, $I_1$, and $I_2$ at this point as the initial conditions.
Depending on the resulting sign
of $\hat \varphi_q(z_+)$, the above middle values become the new ``upper'' or ``lower'' values,
and this bisection procedure continues until 
the uncertainty in $\hat \varphi_q$ at the above point is reduced
to $\Delta_{\rm min}$. Then
we find the next point of the $z$ axis at which the deviation between the ``upper'' and ``lower'' $\hat \varphi_q$-curves
again exceeds $\Delta_{\rm max}$. The cycle continues until the distance between the 
``upper'' and ``lower'' $\hat \varphi_q$-curves at $z=z_+$ does not exceed $\Delta_{\rm max}$.
Obviously, the numbers $\Delta_{\rm min,max}$ are chosen to be sufficiently small. 
Interestingly, the computation time turns out to be quite insensitive 
to $\Delta_{\rm min}$ for a fixed $\Delta_{\rm max}$.

The wake potential is the inverse Fourier transform (over ${\bf k}_\perp$) of the resulting solution $\hat \varphi_q(z, k_\perp)$:
\begin{eqnarray}
\varphi_q ({\bf r}) = \frac{1}{(2\pi)^2}\int \hat{\varphi}_q(z, k_\perp) \exp(i {\bf k}_\perp \cdot {\bf r}_\perp) \, d{{\bf k}_\perp}. \nonumber
\end{eqnarray}
We rewrite the integral in the polar coordinates $k_\perp$, $\alpha$ and integrate analytically over the angle $\alpha$, which yields
\begin{eqnarray}
\varphi_q ({\bf r})
= \frac{1}{2\pi}\int_0^{\infty} d{{k}_\perp} \, k_\perp  \hat{\varphi}_q(z, k_\perp) J_0(k_\perp r_\perp),
\label{inverse-2}
\end{eqnarray}
where $J_0$ is the zero-order Bessel function of the first kind. We note that for sufficiently small $|z|$, 
the integral in Eq.~(\ref{inverse-2}) converges quite slowly at large $k_\perp$, where the numerical integration of the system~(\ref{system}) 
using the method described above turns out to be particularly time-consuming. To circumvent the difficulty, we separate the unscreened 
Coulomb potential from the total potential perturbation. The Fourier transform (over ${\bf r}_\perp$) of the 
Coulomb potential $\varphi_{\rm C}=1/\sqrt{r_\perp^2+z^2}$
is 
\begin{eqnarray}
\hat \varphi_{\rm C} \equiv \int d{{\bf r}_\perp} \, \frac{\exp(-i {\bf k}_\perp \cdot {\bf r}_\perp)}{\sqrt{r_\perp^2+z^2}} \nonumber \\
= \frac{2\pi}{k_\perp} \exp(-k_\perp |z|).
\end{eqnarray}
The potential perturbation $\varphi_q$ can then be written as the sum of $\varphi_{\rm C}$ and the inverse Fourier
transform of $\hat \varphi_q-\hat \varphi_{\rm C}$:
\begin{eqnarray}
\label{inverse-transform-final}
\varphi_q ({\bf r})
= \frac{1}{\sqrt{r_\perp^2+z^2}} + \frac{1}{2\pi}\int_0^{\infty} d{{k}_\perp} \, k_\perp \nonumber \\
\times \left( \hat{\varphi}_q(z, k_\perp) - \frac{2\pi}{k_\perp}\exp(-k_\perp |z|) \right) 
J_0(k_\perp r_\perp).
\end{eqnarray}
The resulting integral converges rather fast. We calculate it using Boole's rule with a sufficiently small fixed step, simply truncating the integration
at a certain sufficiently large $k_\perp$.

\section{Calculation of $\varphi_q$ from Eq.~(\ref{potential-homogeneous})}
\label{calculation-homogeneous}
To calculate the potential perturbation $\varphi_q$ in the homogeneous approximation,
we first rewrite the integral~(\ref{potential-homogeneous}) in the cylindrical coordinates $k_z$, $k_\perp$, $\alpha$
and analytically integrate over the angle $\alpha$. This yields
\begin{eqnarray}
\label{homogeneous-bessel}
\varphi_q({\bf r})=\frac{1}{r} +\frac{1}{\pi} \int_0^\infty dk_\perp \, k_\perp  \int_{-\infty}^{\infty} dk_z \, J_0}{(k_\perp r_\perp)\exp(i k_z z) \nonumber \\
\times \left( \frac{1}{k_\perp^2 + k_z^2 + n_{\rm e 0} - \frac{\displaystyle (k_\perp^2 + k_z^2)n_{\rm i 0}}{\displaystyle (k_z^2 v_{0} -i0^{+})^2}} -\frac{1}{k_\perp^2 + k_z^2} \right),
\end{eqnarray}
where the distances are normalized by $\lambda_{\rm e \infty}$, $n_{\rm e 0}$ by $n_\infty$, $v_0$ by $v_\infty$,
and $\varphi_q$ as per Eq.~(\ref{phi_q-normalization}).

The next step is to analytically perform the integration over $k_z$ in Eq.~(\ref{homogeneous-bessel}) by using the residue 
theorem and Jordan's lemma. 
The poles of the integrand include two real numbers unless the $-i0^{+}$ term is accounted for.
The latter results in infinitesimal {\it positive}
imaginary corrections to both of them. We get for $z>0$:
\begin{eqnarray}
\label{homogeneous-final}
\varphi_q({\bf r})=\frac{1}{r}+\int_0^{\infty} dk_\perp \, k_\perp J_0(k_\perp r_\perp) \nonumber \\
\times 
\left[ \sum_{j=1}^3 S_j(k_\perp, z) - \frac{1}{k_\perp}\exp(-k_\perp z) \right],
\end{eqnarray}
where
\begin{eqnarray}
S_{j}(k_\perp, z)=\frac{i \exp [i k_{zj}(k_\perp) z]}{k_{z j}(k_\perp)+
\frac{\displaystyle n_{\rm i 0} k_\perp^2}{\displaystyle k_{zj}^3(k_\perp) v_0^2}}, \nonumber \\
k_{z 1, z 2}(k_\perp)= \pm \sqrt{B(k_\perp)+C(k_\perp)}, \nonumber \\
k_{z 3}(k_\perp)=i \sqrt{-B(k_\perp)+C(k_\perp)}, \nonumber \\
B(k_\perp)=\frac{n_{\rm i 0}}{2v_0^2}-\frac{k_\perp^2+n_{\rm e 0}}{2}, \nonumber \\
C(k_\perp)=\sqrt{B^2(k_\perp)+\frac{n_{\rm i 0}k_\perp^2}{v_0^2}}.
\end{eqnarray}

We numerically calculate the integral~(\ref{homogeneous-final}) by using Boole's rule with a fixed integration step, truncating
the integration at a sufficiently large $k_\perp$.

\end{document}